\documentclass[aps,pra,twocolumn,showpacs,groupedaddress,amsmath,amssymb]{revtex4}

\usepackage{graphicx}% Include figure files
\usepackage{color}
\usepackage{bm}
\usepackage{latexsym}

\begin{document}

\title{${\cal PT}$-symmetric Talbot Effects}
\author{Hamidreza Ramezani$^1$, D. N. Christodoulides${^2}$, V. Kovanis$^{3}$, I. Vitebskiy$^{3}$, Tsampikos Kottos$^{1,4}$} 
\affiliation{$^1$Department of Physics, Wesleyan University, Middletown, CT-06459, USA}
\affiliation{$^2$College of Optics \& Photonics-CREOL, University of Central Florida, Orlando, Florida 32816, USA}
\affiliation{$^3$Air Force Research Laboratory, Sensors Directorate, Wright Patterson AFB, OH 45433 USA}
\affiliation{$^4$MPI for Dynamics and Self-Organization - Bunsenstra\ss e 10, D-37073 G\"{o}ttingen, Germany}
\date{\today}

\begin{abstract}
We show that complex ${\cal PT}$ -symmetric photonic lattices can lead to a new class of self-imaging Talbot effects. 
For this to occur, we find that the input field pattern, has to respect specific periodicities which are dictated by 
the symmetries of the system. While at the spontaneous ${\cal PT}$-symmetry breaking point, the image revivals occur 
at Talbot lengths governed by the characteristics of the passive lattice, at the exact phase it depends on the gain and 
loss parameter thus allowing one to control the imaging process.
\end{abstract}

\pacs{42.82.Et, 42.25.Bs, 11.30.Er}
\maketitle

%------------------------------------------------------------------------------------------------------
%   Introductory paragraph

{\it Introduction--}The Talbot effect \cite{HFT,LR}, a near field diffraction phenomenon in which self-imaging of a periodic
structure illuminated by a quasi-monochromatic coherent light periodically replicates at certain imaging planes, is an important
phenomenon in optics. These imaging planes are located at even integer multiples of the so-called Talbot distance 
$z_T=2a^2/\lambda$, where $a$ represents the spatial period of the pattern and $\lambda$ the light wavelength. The simplicity 
and beauty of Talbot self-imaging have attracted the interest of many researchers. Such effects find nowadays applications 
in fields ranging from imaging processing and synthesis, photolithography \cite{CEHTSTBWSP}, and optical testing and metrology 
\cite{LSFTB} to spectrometry and optical computing \cite{KP} as well as in electron optics and microscopy \cite{JMC}. Similar 
processes are encountered in other areas of physics involving nonclassical light \cite{KHL}, atom optics \cite{CEHTSTBWSP,JFCSFL}, 
Bose-Einstein condensates \cite{DHDSECHRP}, coupled lasers \cite{PGS01} and waveguide arrays \cite{IMCSMS}. However all these
achievements are limited in studying the properties of the input beams and using {\it real} gratings for imaging. Bypassing 
these limitations will not only enrich the conventional self-imaging research, but also offer new methods for imaging technologies. 
It is therefore extremely desirable to investigate and propose self-imaging architectures which incorporate gain or/and loss mechanisms. 

In the present paper we study the Talbot revivals in a new setting, namely, a class of active lattices with antilinear symmetries. 
These structures deliberately exploits notions of (generalized) parity ($\cal{P}$) and time ($\cal{T}$) symmetry \cite{BB98,BBM99} 
in order to achieve new classes of synthetic meta-materials that can give rise to altogether new physical behavior and novel 
functionality \cite{MGCM08,RMGCSK10,SLZEK11}. Some of these results have been already confirmed and demonstrated in a series of 
recent experimental papers \cite{RMGCSK10,GSDMVASC09,SLZEK11}. In classical optics, ${\cal PT}$-symmetries can be naturally 
incorporated \cite{MGCM08} via a judicious design that involves the combination of delicately balanced amplification and absorption 
regions together with the modulation of the index of refraction. In optics, ${\cal PT}-$symmetry demands that the complex refractive 
index obeys the condition $n({\mathbf r})=n^*(-{\mathbf r})$. It can be shown that these structures have a real propagation constant 
(eigenenergies of the paraxial effective Hamiltonian) for some range (the so-called {\it exact phase}) of the gain and loss 
coefficient. For larger values of this coefficient the system undergoes a {\it spontaneous symmetry breaking}, corresponding to 
a transition from real to complex spectra (the so-called {\it broken phase}). The phase transition point, shows all the characteristics 
of an {\it exceptional point} (EP) singularity. ${\cal PT}$-synthetic matter can exhibit several intriguing features \cite{MGCM08,
GSDMVASC09,RMGCSK10,SLZEK11,K10,RKGC10,ZCFK10,L09,L10d,L10b,CGS11,SXK10,CGCS10,M09,S10,LRKCC11,BFKS09,HRTKVKDC12}. These include 
among others, power oscillations and non-reciprocity of light propagation \cite{MGCM08,RMGCSK10,ZCFK10}, non-reciprocal Bloch 
oscillations \cite{L09}, unidirectional invisibility \cite{LRKCC11} and a new class of conical diffraction \cite{HRTKVKDC12}. In 
the nonlinear domain, such non -reciprocal effects can be used to realize a new generation of optical on-chip isolators and 
circulators \cite{RKGC10}. Other results include the realization of coherent perfect laser-absorber 
\cite{L10b,CGS11} and nonlinear switching structures \cite{SXK10}.

Here, we define conditions which guarantee the existence of Talbot self-imaging for a class of active ${\cal PT}$-symmetric 
lattices. We find that the non-orthogonality of the Floquet-Bloch modes imposed by the non-Hermitian nature of the dynamics 
together with the discreteness of the lattice structures imposes strong constraints for the appearance of Talbot recurrences. 
We show that while at the spontaneous ${\cal PT}$-symmetric point the Talbot length $z_T$ is characterized by the structural 
characteristics of the lattice, in the exact ${\cal PT}$-symmetric phase it is controlled by the gain and loss parameter $\gamma$. 
This allow us to have reconfigurable Talbot lengths for the same initial pattern. Finally, we discuss possible experimental 
realizations where our predictions can be observed.

%--------------------------------------------------------------------------------------------------------------------
\begin{figure}
   \includegraphics[width=.75\linewidth, angle=0]{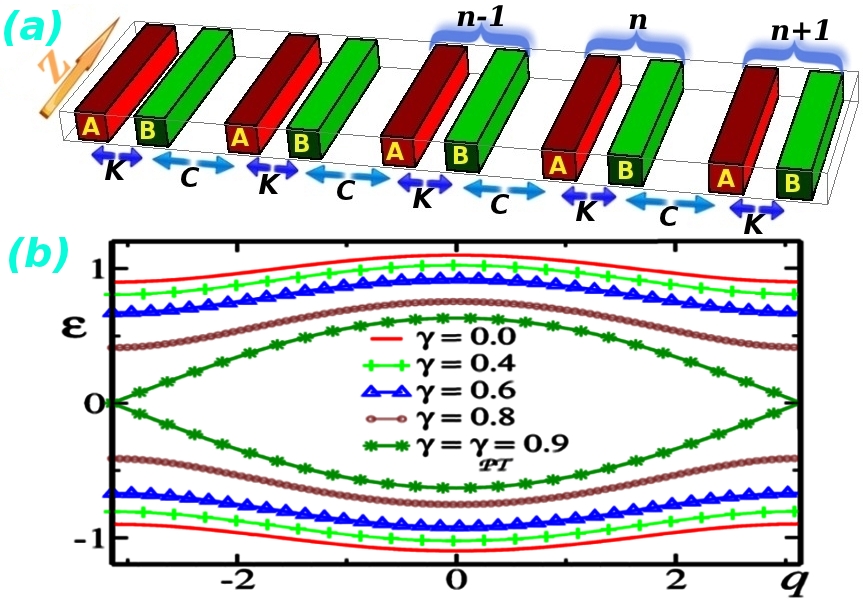}
    \caption{(Color online) (a) Photonic lattice structure with intra-dimer coupling ${k}$ and inter-dimer coupling ${c}$. 
Sublattice (gain waveguide) $A$ is shown by the red rectangular cuboid while sublattice (lossy waveguide) $B$ is shown 
by green rectangular cuboid. Each dimer is distinguished by the index ${n}$. (b) Dispersion relations for various $\gamma$-values.
At $\gamma=\gamma_{\cal PT}$ the gap between the two bands dissappear and an exceptional point singularity is created.
\label{fig:figure1}}
\end{figure}
%--------------------------------------------------------------------------------------------------------------------

\begin{figure}
\includegraphics[width=.75\linewidth, angle=0]{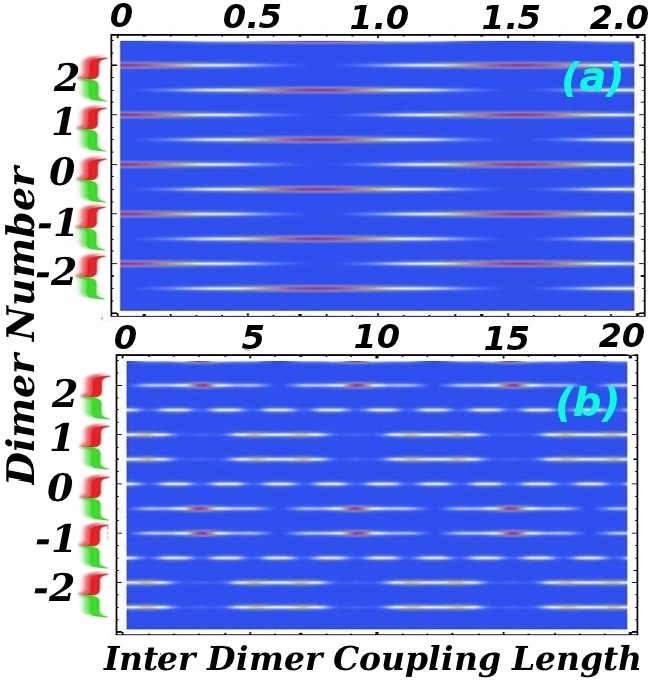}
\caption{(Color online) Talbot intensity "carpets" for period-$N$ input patterns. Length is measured in units 
of inter-dimer coupling $c=1$. The intra dimer coupling is $k=1.05$. (a) periodicity ${N=1}$ with the binary input 
$\{1,0,1,0,...\}$ and $\gamma=0.05$. (b) periodicity ${N=3}$ with the binary input $\{1,1,1,0,0,1,1,1,1,0,0,1,...\}$ 
and $\gamma=\gamma_{\cal PT}=0.05$.
\label{fig2}}
\end{figure}

%==============================================MODEL=========================================================================
{\it Model--} We consider a one-dimensional ($1D$) array of coupled optical waveguides. Each of the waveguides
can support only one mode, while light is transferred from waveguide to waveguide through optical tunneling. The 
array consist of two types of waveguides: type (A) involving a gain material whereas type (B) exhibits an equal 
amount of loss. Their arrangement in space is such that they form $N$ coupled (A-B) dimers with intra and inter-dimer 
couplings $k$ and $c$ respectively, such that both couplings are of similar (but not the same) size i.e. $k\sim c$
(see for example Fig. \ref{fig2} where $k=1.05c$). In the tight binding description \cite{note1}, 
the diffraction dynamics of the electric field amplitude $\Psi_n=(a_n,b_n)^T$ at the $n$-th dimer evolves 
according to the following Schr\"odinger-like equation
\begin{equation}
\label{dyndimer}
\begin{array}{lcr}
i \frac{da_n(z)}{dz} &=& \epsilon a_n(z) + k b_n(z) + c b_{n-1}(z)\\
i \frac{db_n(z)}{dz} &=& \epsilon^{\ast} b_n(z) + k a_n(z) + c a_{n+1}(z)
\end{array}
\end{equation}
where $\epsilon=\epsilon_0+i\gamma$ is related to the complex refractive index \cite{MGCM08}. Without any loss of generality, we will assume below that $\epsilon_0=0$, $\gamma>0$ and $c<k$ \cite{ZCFK10}. The effective Hamiltonian that describes the system commutes with an anti-linear operator (in \cite{ZCFK10} we coined this ${\cal P}_d{\cal T}$-symmetry) which is related with the local ${\cal PT}$-symmetry of each individual dimer.

At this point it is beneficial to adopt a momentum representation $a_n(z)=\frac{1}{2\pi}\int_{-\pi}^{\pi} dq {\tilde a}_q(z)
\exp(inq)$ (and similarly for $b_n$) where the integral is taken over the Brillouin zone $-\pi\leq q \leq \pi$. Because of the translational invariance of the system (\ref{dyndimer}), the equations of motion in the Fourier representation break up into $2\times 2$ blocks, one for each value of momentum $q$:
\begin{equation}
\label{dynfourier}
i\frac{d}{dz}\left(
\begin{array}{c}
{\tilde a}_q(z)\\
{\tilde b}_q(z)
\end{array}
\right)
= H_q\left(
\begin{array}{c}
{\tilde a}_q(z) \\
{\tilde b}_q(z)
\end{array}
\right)
;\quad
H_q=\left(
\begin{array}{cc}
 \epsilon & v_q\\
 v_q^{\ast} & \epsilon^{\ast}
\end{array}
\right)
\end{equation}
with $v_q=k+c\cdot e^{-iq}$. The two component wave functions for different $q$-values are decoupled thus 
allowing for a simple theoretical description of the system. This allows us to perform the evolution in Fourier 
space and then evaluate the spatial representation by a backward transformation i.e.
\begin{equation}
\label{spatialevolved}
  \Psi_n(z)=\frac{1}{2\pi}\int_{-\pi}^{\pi} \psi_q(z)e^{inq}dq.
\end{equation}
where $\Psi_n(z)\equiv (a_n(z), b_n(z))^T$ is the field amplitude for the $n$-th dimer in the spatial representation and 
$\psi_q(z)\equiv ({\tilde a}_q(z), {\tilde b}_q(z))^T$ is the corresponding Fourier component.

%===========================================================================================================
{\it Dynamics --} Substituting in  Eq.~(\ref{dynfourier}) the stationary form $(a_n,b_n)^T=\exp(-i {\cal E}z)(A,B)^T$, 
and requesting non-trivial solutions of the resulting stationary problem, i.e., $(A,B) \neq 0$, we obtain the band 
structure of this diatomic $\cal{PT}$ system \cite{ZCFK10}:
\begin{equation}
\label{dispersion}
{\cal E}_\pm= \pm\sqrt{(k-c)^2+4kc\cos^2(q/2) -\gamma^2}. 
\end{equation}
For $\gamma=0$ we have two bands of width $2c$, centered at ${\cal E} = \pm k$. In this case, the two bands are 
separated by a gap $\delta = 2(k-c)$ and the exact $\cal{PT}$ phase extends over a large $\gamma$ regime. It 
follows from Eq. (\ref{dispersion}) that when $\gamma \geq  \gamma_{\cal PT} = \delta/2$, the gap disappears and 
the two (real) levels at the "inner" band-edges of the two different bands (corresponding to $q= \pm\pi$) become 
degenerate. The corresponding eigenvectors are also degenerate, resulting in an exceptional point (EP) singularity. 
For $\gamma>\gamma_{\cal PT}$ the spectrum becomes partially complex \cite{ZCFK10}. Below we focus our analysis on 
the domain $\gamma\leq \gamma_{\cal PT}$.

The eigenvectors associated with the Hamiltonian Eq. (\ref{dynfourier}) are bi-orthogonal, and therefore do not 
respect the standard (Euclidian) orthonormalization condition. As a result the conservation of total field intensity 
is violated for any $\gamma\neq 0$. Denoting by $|R_{\pm}(q)\rangle=\frac{1}{\sqrt{2}}(1,\frac{{\cal E}_{\pm}(q)-i\gamma}{v_q})^T $ the right eigenvectors corresponding to the eigenvalue ${\cal E}_{\pm}(q)$, we have that the $q-$th momentum components of any initial excitation can be written as $\psi_{q}(0)= \sum_{l=\pm}c_{l}|R_{l}(q)\rangle$. The evolved $q$-field component is 
\begin{equation}
\label{evolved1}
  \psi_{q}(z)=\sum_{l=\pm}c_{l}^qe^{-i{\cal E}_{l}(q)z}|R_{l}(q)\rangle
\end{equation}
where $c_l^q= \langle L_l(q)|\psi_q(0)\rangle$ is the expansion coefficient and $\langle L_l(q)|$ is the left eigenvector
associated with eigenvalue ${\cal E}_{l}(q)$. The above expansion valid as long as the Hamiltonian $H_q$ in Eq. (\ref{dynfourier}) does not have a defective eigenvalue. The latter appears at the spontaneous ${\cal PT}$-symmetric point $\gamma_{\cal PT}=k-c$ (EP) for $q=\pm \pi$. The corresponding evolved $q$-field component is then written as:
\begin{equation}
\label{evolved2}
  \psi_{q=\pm \pi}(z)=(c_1+c_2z)(1, -i)^T +c_2(-i/\gamma , 0)^T
\end{equation}
Direct substitution of Eqs. (\ref{evolved1},\ref{evolved2}) into Eq. (\ref{spatialevolved}) provides the evolution of 
the field in this system. A note of caution is here in order. For the existence of Talbot revivals, a necessary condition is that the initial preparation must not excite the $q=\pm \pi$ defective mode. In the opposite case, the field increases linearly with the propagation distance $z$ (see Eq. (\ref{evolved2})), thus destroying the possibility of revivals of any initial pattern.

%--------------------------------------------------------------------------------------------------------------------
%--------------------------------------------------------------------------------------------------------------------

%===========================================================================================================
{\it Talbot self-imaging --} We are now ready to analyze the Talbot self-imaging recurrences in the case of the ${\cal PT}$-symmetric structure of Fig. \ref{fig:figure1}. We recall that in order the Talbot effect to occur, the input field distribution should be periodic \cite{IMCSMS}, and thus in general $\Psi_n(0)=\Psi_{n+N}(0)$ where $N$ represents the spatial period of the input field. Because of this periodic boundary condition, $q$ can take values only from the discrete set 
\begin{equation}
\label{dis}
  q_m=\frac{2m\pi}{N},\quad m=0,1,2,...,N-1.
\end{equation}
Substituting the above constrain in Eq. (\ref{spatialevolved}) we get the following expression for the evolved field at the 
$n$-th dimer
\begin{equation}
\label{evolvedm}
  \Psi_{n}^{(N)}(z)=\sum_{l=\pm;m=1}^{N-1}c_{l}^{q_m}e^{-i{\cal E}_{l}(q_m)z}|R_{l}(q_m)\rangle
\end{equation}
It is therefore clear that field revivals are possible at intervals $z$ if ${\cal E}(q_m)z_T=2\pi\nu$ where $\nu$ is an integer. Therefore the ratio of any two eigenvalues ${\cal E}_m\equiv {\cal E}(q_m)$ has to be a rational number, i.e.
\begin{equation}
\label{condition}
  \frac{\sqrt{(k-c)^2+4kc\cos^2(\frac{m\pi}{N}) -\gamma^2}}{\sqrt{(k-c)^2+4kc\cos^2(\frac{m'\pi}{N}) -\gamma^2}}=\frac{\alpha}{\beta}
\end{equation}
where $\alpha$ and $\beta$ are relatively prime integers. At the same time, revivals in the field intensity are ensured
provided that $({\cal E}_m-{\cal E}_{\mu})/({\cal E}_{m'}-{\cal E}_{\mu'})=\alpha/\beta$ where the indices belong to the 
set $\{0,1,...,N-1\}$ and are taken at least three at a time. It is straightforward to show that this condition is trivially satisfied for the same set of $N$-input pattern periodicities as for the fields. 

Next we consider the field Talbot revivals of input patterns with period $N$, at the spontaneous ${\cal PT}$-symmetric point. To this end, we observe that a direct substitution of $\gamma=\gamma_{\cal PT}$ in Eq. (\ref{condition}) for the ratio ${\cal E}_m/{\cal E}_0$ leads to the simple condition $\cos(m\pi/N)= \alpha/\beta$. The latter is re-written in terms of the Chebyshev polynomials which are defined as $\cos(mx)= T_m(\cos(x))=\sum_{j=0}^{[m]} c_j^{(m)}(\cos(x))^{m-2j}$, where $[m]$ represents the integer part of $m$. The Chebyshev coefficients $c_j^{(m)}$ are integer numbers and, of importance to our discussion, is the fact that the first one is given by $c_0^{(m)} =2^{m-1}$. Given that $c_j^{(m)}$ are integers, then $\cos(\frac{\pi m}{N})$ is rational if and only if $\cos(\frac{\pi}{N})$ is rational \cite{IMCSMS}. Using the Chebyshev identity with $m=N$ (assuming $N$ is an odd number), we obtain the following polynomial in $\cos(\frac{\pi}{N})$:
\begin{equation}
  \label{poly}
2^{N-1}(\cos(\frac{\pi}{N}))^{N}+...+c_{[N/2]}^{N} \cos(\frac{\pi}{N})+1=0
\end{equation}
where we have used the fact that $T_{N}(\cos{\frac{\pi}{N}}) = \cos(N\pi/N)=-1$. By applying the rational root 
theorem one can show that the roots of this polynomial in $\cos(\pi/N)$ are rational only if $N=1,3$. Similar 
techniques leads to the fact that for even values of $N$ the only possibility is $N=2$ \cite{IMCSMS}. However, 
input patterns with $N=2$ periodicity, excite the $q=\pm\pi$ Fourier mode, and therefore based on our previous 
discussion (see Eq.~(\ref{evolved2}) above), have to be excluded. Therefore, strictly speaking, discrete Talbot 
revivals at the spontaneous ${\cal PT}$-symmetric point are possible only for a finite set of periodicities $N=1,3$, 
where for example, the $N=1$ case can represent initial patterns $\{1,0,1,0\cdots,1,0\}$ or $\{0,1,0,1\cdots,0,1\}$ 
or the more trivial case of a plane wave with $\{1,1,1,1\cdots,1,1\}$. Some representative intensity revivals for 
$N=1$ and $3$ periods are depicted in Fig. \ref{fig2}.

%--------------------------------------------------------------------------------------------------------------------
\begin{figure}
   \includegraphics[width=1\linewidth, angle=0]{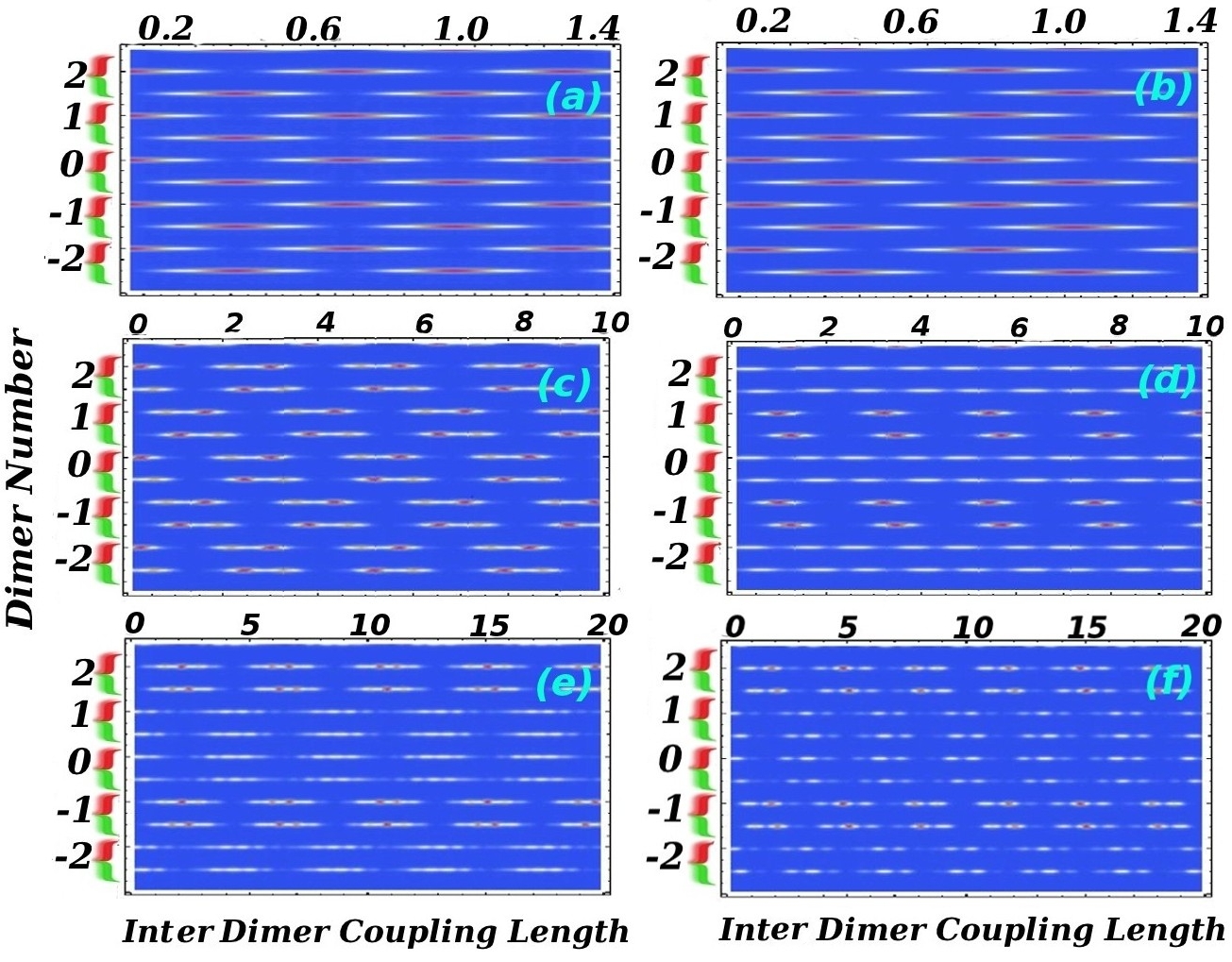}
    \caption{(Color online) Talbot intensity "carpets" for period-$N$ input field patterns at the exact phase $\gamma<\gamma_{\cal PT}$. Everything is measured in units of inter-dimer coupling $c=1$ while the intra-dimer coupling is $k=4$. (a) $\gamma=0.1$ while in (b) $\gamma=2$. In both cases the input pattern has periodicity $N=1$ and it is chosen to be $\{1,0,1,0,1,0,1,0,\cdots \}$; (c) $\gamma=\sqrt{11/3}$ and (d) $\gamma=\sqrt{7}$. Now the input pattern has periodicity $N=2$ and it is chosen to be $\{1,1,0,0,1,1,\cdots \}$; (e) $\gamma=5/\sqrt{17}$ and (f) $\gamma=9/\sqrt{10}$. The input pattern in these cases is $\{1,1,1,1,0,0,1,1,1,1,0,0\cdots \}$ and has periodicity $N=3$. Different Talbot lengths $z_T$ are observed between (a-b) and (c-d) and (e-f). 
\label{fig3}}
\end{figure}

%--------------------------------------------------------------------------------------------------------------------

The Talbot revivals can appear also in the exact phase $\gamma<\gamma_{\cal PT}$. Simple inspection of Eq. (\ref{condition})
indicates that an initial periodic pattern with periodicity $N=1$ (resulting to eigenvalue index $m=0$ in Eq. (\ref{dis}))
leads to a rational value $\alpha/\beta=1$. In this case the Talbot length $z_T$ depends on the gain and loss parameter as
$z_T=2\pi/{\cal E}_0=2\pi/\sqrt{\gamma_{\cal PT}^2+4kc-\gamma^2}$ and therefore it varies by changing $\gamma$. Such
reconfigurable behavior of the Talbot length is characteristic of the exact phase $\gamma<\gamma_{\cal PT}$ and can be 
found also for the $N=2,3$-period input patterns. For $N=2$ (corresponding to eigenvalue indices $m=0,1$ in Eq. (\ref{dis})) 
one can show that for fixed $k,c$ and $\gamma_{\cal PT}= k-c$ such that $\gamma_{\cal PT}/(k+c) > \alpha/\beta$, Eq. (
\ref{condition}) is satisfied provided that $\gamma=\sqrt{\gamma_{\cal PT}^2-4kc\alpha^2/(\beta^2-\alpha^2)}$ (we assume 
that $\alpha<\beta$). Similarly for $N=3$, the Talbot revivals are possible provided that $\gamma=\sqrt{\gamma_{\cal PT}^2+
kc[1-4(\alpha/\beta)^2]/[1-(\alpha/\beta)^2]}$ where $0.5<\alpha/\beta<\sqrt{1-3kc/(k+c)^2}$. In both cases the corresponding 
Talbot length is $\gamma$-dependent and it is given by the largest period $z_T=2\pi/|{\cal E}_j-{\cal E}_l|\sim{2\pi/{\cal E}_0}$ 
that results from the eigenvalues involved in the initial pattern. Example cases of Talbot self-imaging revivals for initial 
periodic patterns with period $N=1,2,3$ and different $\gamma$ values are shown in Figs.~\ref{fig3}a-b; Figs.~\ref{fig3}c-d; 
and Figs.~\ref{fig3}e-f respectively. We see that for the same initial preparation, the revivals are controlled by $\gamma$ 
and can occur at different Talbot lengths.

In fact, we can show that larger periods $N>3$ do not result in Talbot self-imaging revivals in the exact ${\cal PT}$-symmetric
domain. Using Eq.(\ref{condition}), for the $ |{\cal E}_m|/|{\cal E}_0|=\alpha/\beta$ and enforcing the constrain that
$\gamma\leq\gamma_{\cal PT}=k-c$, one obtains the inequality, $\cos(\frac{m\pi}{N})\leq \frac{\alpha}{\beta}$ which has
to be satisfied together with the equation Eq. (\ref{condition}) (the equality correspond to the case $\gamma=\gamma_{\cal PT}$ 
discussed above). At the same time $\cos(\frac{m\pi}{N})$ has $m=0,\cdots,N-1$ roots. By applying the intermediate value theorem 
one finds out that this inequality cannot be valid for $N>3$. 

%============================================================================================================================
{\it Experimental Implementation--} We would like finally to suggest possible experimental implementations of the ${\cal PT}$-
symmetric waveguide arrays, which will allow for the observation of the reconfigurable Talbot effect. The proposed structures 
will involve MBE grown quantum wells (QW) that will be patterned to form coupled waveguides. The basic ${\cal PT}$ structural 
element of the array shown in Fig. \ref{fig:figure1}, involves two ${\cal PT}$-symmetric sites (dimer). Such a design is desirable 
because of its simplicity. The dimensions and index contrast can be such that each waveguide will be single-moded. For example, 
for AlGaAs structures this can be achieved by a refractive index of $n_0 =3.35$ operated at 800 nm.
Reconfigurable gain can be achieved by running an electric current through a AlGaAs/GaAs QW p-n junction. In such structures one 
can easily reach gain and loss values as high as $50cm^{-1}$. The two site channels in every dimer will be excited at different current 
levels $I_1$ and $I_2$ so as to establish the antisymmetric gain and loss profile that is necessary to observe $\cal PT$ optical behavior. In 
practice this will be done provided that current $I_1 \gg I_2$ so as the corresponding regions underneath see equal amount of gain 
and loss. More specifically, $I_2$ will be relatively small so the associated waveguide site will experience material absorption. 
Its sole purpose will be for fine tuning. Given that $I_1$ and $I_2$ can be interchanged and adjusted, this will allow us to dynamically 
control the Talbot length $z_T$ of these ${\cal PT}$-symmetric structures. Of course, special consideration has to be given to the 
effects of gain and loss on the modal index change in these structures (because of the Kramers-Kronig relations). 

Finally we comment on the robustness of Talbot revivals against structural imperfections. For realistic values of positional 
imperfections (up to 5\% of the inter-dimer coupling) we have confirmed numerically that Talbot revivals are only slightly 
distorted. Specifically we found that revivals associated with short Talbot lengths $z_T$, are {\it essentially} unaffected for 
moderate propagation distances $z$ while revivals associated with larger lengths $z_T$ are fragile due to the distortion of 
the delicate balance between the mode amplitudes and phases that eventually dominate the evolution.

{\it Conclusions--} In conclusion, we have shown that a class of ${\cal PT}$-symmetric optical lattices, support Talbot
self-imaging revivals for input patterns with periodicities dictated by the discreteness of the lattice and the strength of 
gain and loss parameter. Of interest will be to investigate if Talbot revivals can also occur in higher dimensions and in the presence 
of non-linearity. Our results might be applicable to other areas like self-imaging of coupled lasers \cite{PGS01} with 
distributed gain and synchronization of ${\cal PT}$-symmetric coupled electronic oscillators \cite{SLZEK11}. 

{\it Acknowledgments --} 
This research was supported by AFOSR via grants No. FA 9550-10-1-0433, No. FA 9550-10-1-0561 and LRIR 09RY04COR, and by an NSF ECCS-1128571 grant. Vassilios Kovanis and Ilya Vitebskiy  were supported via the Electromagnetics Portfolio of Dr. Arje Nachman of AFOSR.
%-----------------------------------------------------------------------------------------------------------------------------

%%%%%%%%%%%%%%%%%%%%%%%%%%%%%%%%%%%%%%%%%%%%%%%%%%%%%%%%%%%%%%%%%%%%%%%%%%%%%%%%%%%%%%%%%%%%%%%%%%%%%%%%%%%%%%%%%%%%%%%%%%%%%%%%%%%%%%


\begin{thebibliography}{99}
\bibitem{note1} This approximation, holds if the waveguides that the array is made of are single-mode, exhibit a strong
confinement and a weak overlap with the modes of the adjacent guides, thus entailing only nearest neighbor interactions
(see for example discussion at \cite{L08}).
\end{thebibliography}
\end{document}